\documentclass[12pt]{article}
\usepackage{cite}
\makeatletter
\@addtoreset{equation}{section}
\makeatother

\newcommand{\be}[3]{\begin{equation}  \label{#1#2#3}}

\newcommand{\ee}{ \end{equation}}
\newcommand{\ba}{\begin{array}}
\newcommand{\ea}{\end{array}}
\newcommand{\bea}{\begin{eqnarray}}
\newcommand{\eea}{\end{eqnarray}}
\newcommand{\p}{\partial}
\newcommand{\z}{\phi}

\newcommand{\remark}[1]{}
\let\LARGE=\Large
\let\Large=\large
\begin{document}

%%%%%%%%%%%%%%%%%%%%%%%%%%%%%%%%%%%%%%%%%%%%%%%%%%%%%%%%

\thispagestyle{empty}
\rightline{CALT-68-2294}
\rightline{CITUSC/00-046}
\rightline{hep-th/0008112}

\vspace{8truemm}

\centerline{\bf \LARGE 
Domain walls in five dimensional supergravity}

\bigskip

\centerline{\bf \LARGE 
with non-trivial hypermultiplets}
\bigskip

\vspace{1truecm}

\centerline{\bf Klaus Behrndt$^a$\footnote{e-mail: 
 behrndt@theory.caltech.edu}, \ 
Carl Herrmann$^b$\footnote{e-mail: 
 herrmann@hera1.physik.uni-halle.de}, \ 
Jan Louis$^b$\footnote{e-mail: 
 j.louis@physik.uni-halle.de}\ {\rm and}\ 
Steven Thomas$^c$\footnote{e-mail: 
 S.Thomas@qmw.ac.uk}}

\vspace{.5truecm}
\centerline{$^a$ \em California Institute of Technology}
\centerline{\em Pasadena, CA 91125, USA}

\medskip

\centerline{\it CIT-USC Center For Theoretical Physics}
\centerline{\it University of Southern California}
\centerline{\it Los Angeles, CA 90089-2536, USA}

\vspace{.4truecm}

\centerline{$^b$ \em Fachbereich Physik, Martin-Luther-Universit\"at
Halle-Wittenberg}
\centerline{\em Friedemann-Bach-Platz 6, D-06099 Halle, Germany}

\vspace{.4truecm}

\centerline{$^c$ \em 
Department of Physics, Queen Mary and Westfield College}
\centerline{\em Mile End Road, London E1, U.K.}

\vspace{1truecm}

%%%%%%%%%%%%%%%%%%%%%%%%%%%%%%%%%%%%%%%%%%%%%%%%%%%%%%%%

\begin{abstract}
We study BPS domain wall solutions of 5-dimensional $N=2$ supergravity
where isometries of the hypermultiplet geometry have been gauged.  We
derive the corresponding supersymmetric flow equations and define an
appropriate c-function.  As an example we discuss a domain wall
solution of Freedman, Gubser, Pilch and Warner which is related to a
RG-flow in a dual superconformal field theory.

\end{abstract}

August 2000

%\today

%%%%%%%%%%%%%%%%%%%%%%%%%%%%%%%%%%%%%%%%%%%%%%%%%%%%%%%%%%%%%%%%

\newpage

%%%%%%%%%%%%%%%%%%%%%%%%%%%%%%%%%%%%%%%%%%%%%%%%%%%%%%%%%%%%%%%%%

\section{Introduction}

%%%%%%%%%%%%%%%%%%%%%%%%%%%%%%%%%%%%%%%%%%%%%%%%%%%%%%%

The AdS/CFT correspondence (for a review see \cite{401} and references
therein) relates gauged supergravity in a background geometry
$AdS_5\times H$ to a superconformal field theory (SCFT) in a
four-dimensional Minkowski space living on the the boundary of
$AdS_5$.  The Kaluza-Klein excitations of the supergravity are
identified with gauge invariant operators in the SCFT.  $H$ is called
the horizon manifold and its isometries are related to the R-symmetry
of the superconformal algebra.  The original example of \cite{402}
considered $H=S^5$ whose isometry group $SO(6)$ corresponds to the
R-symmetry of the dual $N=4$ SCFT. Examples with less supersymmetry
and different $H$-manifolds have been constructed for example in
\cite{403,411,404,456,405}.

It is possible to perturb the SCFT by adding an
operator $O_h$ to the action
\be001
S \ \to \ S + h\int d^4x\, O_h(x) \ .
\ee
In general this breaks the conformal symmetry and if the perturbation
is relevant a renormalization group (RG) flow for the coupling $h$ is
induced.  The resulting infrared (IR) theory can be free, confining or
have a non-trivial fixed point where the $\beta$-function vanishes.
In the latter case the RG-flow connects two different CFT, the
original ultra-violet (UV) theory with the IR theory.

In the dual supergravity description the coupling $h$ is identified
with a scalar field $\Phi$ and the RG-flow corresponds to a domain
wall (DW) solution which interpolates between two different extrema of
the scalar potential $V(\Phi)$ \cite{351,352,041,040,120,121}.  If the
UV-theory and the IR-theory are both conformal the two extrema
necessarily have to be $AdS_5$ vacua of $V$.  The DW solution requires
that the scalar field $\Phi$ has a non-trivial dependence on the the
radial coordinate of $AdS_5$ which can be identified with the energy
scale $\mu$ of the RG-flow \cite{120,121,406,407}.

A particular example of such an RG-flow has been presented in
\cite{040} and further discussed in \cite{410,408}.  It is a flow from
an $N=4$ SCFT in the UV to an $N=1$ SCFT in the IR which preserves
$N=1$ supersymmetry along the flow.
In the dual supergravity description this was identified with a DW
solution of five-dimensional gauged $N=8$ supergravity connecting two
$AdS_5$ vacua.  The solution preserves four supercharges (it is a BPS
solution) and interpolates between two extrema one of which preserves
the full $N=8$ supersymmetry and is identified with the UV SCFT while
the second extremum only preserves $N=2$ supersymmetry and is
identified with the IR SCFT.  The BPS property corresponds to the fact
that $N=1$ supersymmetry is preserved along the RG-flow.

The purpose of this paper is to study BPS domain wall solutions of
five-dimensional $N=2$ gauged supergravity which preserve half of the
supercharges ($N=1$).\footnote{We use the convention of $D=4$ to count
supercharges. The minimal supersymmetry in $D=5$ has 8 supercharges
which we call $N=2$ throughout this paper.}  This establishes the
framework for generalized RG-flows which start from an UV theory with
less ($N=1,2$) supersymmetry.  It also simplifies the analysis since
the $N=2$ scalar potential is somewhat less involved than its $N=8$
counterpart.

An additional motivation arises from the fact that such DW solutions
are closely related to a supersymmetric version of the Randall-Sundrum
scenario \cite{050,271,272,273,274,275,320}.  In this scenario gravity is
localized near the wall through exponential suppression and therefore
requires a DW which asymptotes to IR fixed points on both sides.  It
has been shown in \cite{090,110} that there are no such fixed points
for theories containing vector/tensor multiplets, but we will argue
that this does not necessarily apply if charged hypermultiplets are
present.

BPS domain wall solutions of five-dimensional $N=2$ supergravity have
been studied previously.  In refs.\ \cite{190,250} the DW solutions
arising from compactification of 11-dimensional Hor\v ava--Witten
M-theory on Calabi-Yau threefolds were derived. In this case the
necessity of non-trivial four-form flux results in the gauging of an
axionic shift symmetry which is an isometry of the universal
hypermultiplet.  Refs.\ \cite{090,020,010} considered DW solution with
non-trivial vector multiplets and showed that within this setup no IR
fixed point can arise.  As an immediate corollary also supersymmetric
RS domain walls can not be obtained with only vector multiplets
\cite{090}.  Non-trivial tensor multiplets were considered in refs.\
\cite{220,210} but this does not alter the conclusion about possible
IR fixed points.  Finally ref.\ \cite{221} derived the
five-dimensional gauged $N=2$ supergravity including vector-, tensor-
and hypermultiplets. 
Many aspects of these discussions go in parallel to domain walls
in 4-dimensional gauged supergravity \cite{201, 200}.

In this paper we consider both vector- and hypermultiplets and derive
the condition for a BPS domain wall solution including charged
hypermultiplets.  Such a solution is only possible 
for Abelian gauge symmetries.  
We argue that in this case the previous
`no-go' theorems do not apply.  Specifically, in section 2 we recall a
few facts about $N=2$ gauged supergravity with particular emphasis on
gauged isometries in the hypermultiplet sector.  It turns out that we
have to allow for more general gaugings of the $SU(2)_R$ symmetry than
have previously been considered.  In section 3 we study BPS domain
wall solutions with both non-trivial vector- and hypermultiplets.  We
derive the supersymmetric flow equations for the scalar fields and
show that the corresponding c-theorem is satisfied.  As an application
of the formalism we discuss in section 4 the RG-flow of ref.\
\cite{040}.  Section~5 presents our conclusions and contains a
preliminary discussion of a smooth supersymmetric RS domain wall.

%%%%%%%%%%%%%%%%%%%%%%%%%%%%%%%%%%%%%%%%%%%%%%%%%%%%%

\section{$N=2$ gauged supergravity}

%%%%%%%%%%%%%%%%%%%%%%%%%%%%%%%%%%%%%%%%%%%%%%%%%%%%%
A generic spectrum of five-dimensional $N=2$ supergravity contains the
gravitational multiplet, $n_v$ vector multiplets in the adjoint
representation of some gauge group $G$, $n_h$ hypermultiplets which
can be charged under $G$ and tensor multiplets.  The gravitational
multiplet contains the graviton $g_{mn}$, two gravitini $\psi_m^A,
A=1,2$ which are symplectic-Majorana spinors\footnote{For conventions
and notations see \cite{190,200}.} and the graviphoton $A_m^0$.  A
vector multiplet contains a vector $A_m$, two gaugini $\lambda^A$ and
a real scalar $\phi$ while the hypermultiplet features two hyperini
$\zeta^\alpha$ and four real scalars $q^u$.

A tensor multiplet has the same field content as a vector multiplet
but with the vector replaced by a tensor. In $D=5$ vector and tensor
are dual to each other and this duality can be performed as long as
the tensor fields are neutral under $G$ \cite{220,210,409}.  In this
paper we consider this case and thus only keep $n_v$ vector and $n_h$
hypermultiplets in the spectrum.

The action for scalar fields coupled
to supergravity is given by
\be020
S \ = \ \int d^5 x\ \sqrt{-g}
\Big[{\frac{1}{2}} R - V(\Phi) - {\frac{1}{2}} 
g_{MN} \partial_{m} \Phi^M \partial^m \Phi^N
\Big] ,
\ee
where we have omitted gauge fields and fermions.  The $\Phi^M$
collectively denote the scalar fields $\z^i, i=1,\ldots, n_v$ in the
vector multiplets and the scalar fields $q^u, u=1,\ldots,4n_h$ in the
hypermultiplets, i.e.\ $\Phi^M =(\z^i, q^u)$.  Supersymmetry forces
the metric to be block diagonal
\be021
g_{MN} \ = \ \left(\ba{cc} g_{ij}&0\\
                      0& g_{uv}\ea\right)\ ,
\ee 
where the metric for the vector
multiplets $(g_{ij})$ has to be very special K\"ahler 
\cite{080} and the metric
for the hypermultiplets $(g_{uv})$ has to be 
quaternionic.

The very special K\"ahler geometry is best described by introducing
$n_v+1$ functions $X^I(\z^i), I=0,\ldots,n_v$ which satisfy one
contraint equation
\be123
{\cal V}\ \equiv \
\frac16 C_{IJK}X^IX^JX^K \ = \ 1\ ,
\ee
where the $C_{IJK}$ are arbitrary constants
determining the scalar manifold.
The metric $g_{ij}$ is then obtained via
\be999
g_{ij} \ = \ \partial_iX^I\partial_jX^J\,G_{IJ}|_{{\cal V}=1}\ ,
\qquad 
G_{IJ} \ = \ -\frac12\,\partial_I\partial_J
\ln {\cal V} |_{{\cal V}=1}\ .
\ee

The $4n_h$ scalars in the hypermultiplets are
coordinates on a quaternionic manifold \cite{200,221}. 
This implies the existence of three (almost) complex 
structures $(J^x)^w_v, x=1,2,3$ which satisfy the
quaternionic algebra.
Associated with the complex structures is a 
triplet of K\"ahler forms 
$K^x_{uv} = g_{uw} (J^x)^w_v$
where $g_{uw}$ is the quaternionic metric.
The holonomy group of a quaternionic manifold
is $Sp(2)\times Sp(2n_h)$ and 
$K^x$ is identified with the field strength of the
$Sp(2)\sim SU(2)$ connection $\omega^x_v$ i.e.
\be653
K^x \ = \
d\omega^x +\frac12\,\epsilon^{xyz}\omega^y\omega^z~.
\ee
As a consequence the $K^x$ are covariantly closed with respect to the $SU(2)$
connection $\omega^x$
\be998
dK^x + \epsilon^{xyz}\omega^yK^z\ = \ 0~.
\ee
For later use we need to introduce the 
quaternionic vielbeins $V_u^{A\alpha}$
defined via
\be993
 g_{uw} \ = \ V_u^{A\alpha} V_w^{B\beta} \epsilon_{AB} 
C_{\alpha\beta}\ ,
\ee
where $\alpha,\beta =1,\ldots, 2n_h$ and
$C_{\alpha\beta}$ is the flat $Sp(2n_h)$ metric.

The isometries on the scalar manifold are generated
by a set of $n_v+1$ 
Killing vectors $k_I^u(q),\, k_I^i(\phi)$
\be995
\delta q^u \ = \ \epsilon^I k_I^u(q) \ ,\qquad
\delta \phi^i \ = \ \epsilon^I k_I^i(\phi) \ .
\ee
The $k_I^u$ are determined by a 
triplet of Killing prepotentials $P_I^{x}(q)\,$ 
via\footnote{For $SU(2)$ 
valued matrices, we  adopt the convention 
$Y^A_{\ B} =  iY^x\,(\sigma^x)^A_{\ B}$.}
\be996
k^u_I\,K_{uvB}^A \ = \  \partial_v P_{IB}^A 
+ [\omega_{v},P_I]_{\ B}^A\ .
\ee
Furthermore, the $P_I$ satisfy a Poisson bracket relation~:
\be678
\{P_I,P_J\}^A_{\ B} \ \equiv \
k^u_Ik^v_J\, K_{uvB}^A -[P_I,P_J]^A_{\ B} \ = \
-\frac12\,f_{IJ}^K\,P^A_{K\,B}~,
\ee
where $f_{IJ}^K\,$ is the structure constant of the isometry algebra.
It is possible to gauge (part of) these isometries
by modifying the covariant derivatives of the scalars,
gaugini and hyperini, 
their supersymmetry transformation and the Lagrangian. Furthermore it is 
possible to (independently)
gauge the $SU(2)_R$ symmetry of the $N=2$ supergravity. 
This modifies the covariant derivatives of the gravitino and the
gaugino but not the hyperino. Finally, it is also possible
to simultaneously gauge the isometries and the R-symmetry. 
We do not recall the details of this somewhat technical enterprise here 
but refer the reader to 
the literature \cite{200,190,221}.

However, for the purpose of this paper, we are lead to consider more
general gaugings of the R-symmetry. We modify the procedure outlined
in \cite{221}, in that we shift the space-time pullback of the $SU(2)$
connection by the following linear combination of the vector fields
\be030
\omega_{\ B}^A \ \to \ \widehat\omega_{\ B}^A \ \equiv \ 
\omega_{\ B}^A + A^I \widehat P_{I B}^A(q)\ ,
\ee
 where $ \widehat P_{IB}^A$ is an $SU(2)$ valued matrix.
This differs from the procedure of ref.\ \cite{221} by the choice of the
matrix $ \widehat P_{IB}^A$. In \cite{221} the Killing prepotential was used
i.e.\ 
$\widehat P_{IB}^A = P_{IB}^A $ 
while we allow for 
the possibility of having a slightly more general
$\widehat P_{IB}^A$ which differs from $P_I$
by a $q$-dependent scalar (i.e.\ $SU(2)$ invariant) function
\be031
\widehat P_{IB}^A \ = \ \widehat{\gamma}_I(q)\, P_{IB}^A\ 
\quad \mbox{(no sum on $I$)}.
\ee
A rigorous proof
 of the consistency of this ansatz is beyond the scope of this
paper and we leave it for future investigation. However, as a first check 
we verified that the new $SU(2)$ curvature $\hat K$ defined analogously to
(\ref{653}) with $\omega$ replaced with $\widehat\omega$ satisfies 
\be669
\hat K^A_{\ B} \ = \ K_{uvB}^A
{\cal D}q^u{\cal D}q^v + \widehat F^I
P^A_{I\,B}~.
\ee
Eq.\ (\ref{669}) holds in virtue of
(\ref{996}) and (\ref{678}), 
with the definitions
\be387
{\cal D}q^u \ = \ dq^u -\frac12\widehat A^Ik_I^u~, \quad \widehat F^I =
d\widehat A^I+\frac14 f_{KL}^I\widehat A^K\widehat A^L~,\quad 
\widehat A^I = \widehat\gamma^I(q) A^I~.
\ee
$\hat K$ still satisfies eq.\ (\ref{998}) with $\omega$
replaced with $\widehat\omega$.

Thus the only modification following from (\ref{030}) is to replace
the gauge fields $A^I$ and their field strength with their hatted
counterparts in terms arising from the R-symmetry gauging. As these
terms are supersymmetric by themselves \cite{221}, this suggests the
consistency of our ansatz.\footnote{We thank A. Ceresole and
G. Dall'Agata for discussions on this point.} 
For $\widehat P_I = P_I$ the consistency has been shown in \cite{221} while
for $\widehat P_I \neq P_I$ this has not been firmly
established yet.  The interpretation of the field dependent vector fields
$\hat A^I$ in the light of the AdS/CFT correspondence will be given at
the end of section 3.

Finally, gauging the isometries and the R-symmetry also
requires the presence of a scalar potential
\cite{200,190}. Using the modified $SU(2)$ connection, one obtains
\be997
V \ = \ (-2 G^{IJ}+4\,X^IX^J)\,
Tr(\widehat P_I\widehat P_J)
+ X^IX^J (g_{uv}\,k^u_Ik^v_J 
+g_{ij}\, k_I^i k_J^j)\ ,
\ee
where the first term can be traced back to the gauging of the $SU(2)_R$
symmetry and the second term to the gauging of the isometries of the scalar
manifold. For $\widehat P_I=P_I$, this potential coincides with the
potential of refs.\ \cite{200,190}. 

%%%%%%%%%%%%%%%%%%%%%%%%%%%%%%%%%%%%%%
\section{BPS domain wall solutions}
%%%%%%%%%%%%%%%%%%%%%%%%%%%%%%%%%%%%%%
%
In this section we derive BPS domain wall solutions which preserve
half of the eight supercharges.  As an Ansatz for a
metric which respects 4-d Poincare invariance we use 
\be010
ds^2 \ \equiv \ g_{mn}dx^m\,dx^n\ = \ \mu^2 ( -dt^2 + d\vec x^2) 
%+(\mu\, W(\mu))^{-2} d\mu^2
+  \frac{d\mu^2}{\mu^2\,W^2(\mu)} \quad .
\ee
For constant $W$ this is the metric of $AdS_5$.  In the AdS/CFT
correspondence the fifth coordinate $\mu$ will be identified with an
energy scale in the dual four-dimensional field 
theory \cite{120,121,406,407}.  The UV region
(= large length scale in supergravity) corresponds to $\mu \rightarrow
\infty$, while the IR is approached for $\mu \rightarrow 0$.  For
later use we also record that for this metric the vielbeins and the
non--vanishing spin connections are given by
\be337
e_t{}^0\ = \ e_x{}^a\ = \ \mu~,\qquad e_\mu{}^4 \ = 
\ (\mu\,W)^{-1}\ ,\qquad
 \omega_{t\,04}\ = \ \omega_{x\,a4} \ = \ \mu\,W\ ,
\ee
where $(0,a,4)$ are the tangent space indices.

We require that the DW solutions preserves four supercharges and as a
consequence we have to demand that the supersymmetry variations of the
two gravitini $\psi_\mu^A, A=1,2$, the $2n_v$ gaugini $\lambda^{A}_i$
and the $2n_h$ hyperini $\zeta^\alpha$ admit four Killing spinors in
the bosonic background specified by the metric (\ref{010}) and
vanishing gauge fields. Specifically we demand
\begin{eqnarray}\label{susyvar1}
\delta\psi_m^{A} &=& D_m\epsilon^{A}
-\frac{i}{3}\,X^I\,\widehat P^A_{IB}\,\Gamma_m\epsilon^B \ = \
0\ ,
\\ \label{susyvar2}
\delta \lambda^{Ai} &=& -\frac{i}{2} 
\Gamma^m \partial_m \phi^i\,\epsilon^{A}
+ g^{ij}(\partial_j X^I)\,
\widehat P^A_{IB}\,\epsilon^B\ +  X^I k_I^i
\,\epsilon^A \ = \ 0\ ,
\\\label{susyvar3}
\delta\zeta^\alpha &=& 
-\frac{i}{\sqrt{2}}\, V^{A\alpha}_u\,\Big( \Gamma^m \partial_m q^u\, 
+ i X^I k_I^u \Big) \epsilon_A \ = \ 0 \ .
\end{eqnarray}
We are mainly interested in the dependence of the scalar fields on the
fifth coordinate or, in terms of the dual field theory, on the energy
scale $\mu$. Hence we ignore the 4-d spacetime dependence and consider
the scalars $\Phi(\mu)$ only as functions of $\mu$.

Let us first consider 
the $(t,x)$ components of eq.\ (\ref{susyvar1}). 
They imply a projection on the supersymmetry
parameters $\epsilon^B$
\be012
\Big(W\,\Gamma^4 \delta_B^A
-\frac{2i}{3}\, 
X^I \widehat P^A_{I B} \Big)\epsilon^B \
= \ 0~.
\ee
For this to be a consistent projector one learns from 
$(\Gamma^4)^2 = 1$  that 
\be013
W^2\,\delta^A_{\ B} \ = \ -\frac{4}{9}\, 
(X^I\widehat P_I X^J\widehat P_J)^A_{\ B}~.
\ee
An additional constraint arises 
if one demands that (\ref{012}) admits
four Killing spinors.
This is the case if all $SU(2)$ matrices $\widehat P_I$ can
simultaneously be rotated in the direction of $\sigma ^3$ or
equivalently if
\be310
[\widehat P_I,\widehat P_J]\ =\ 0
\ee
holds.\footnote{One could consider $X^I\widehat P_I$ as a
single $q$ and $\phi$ dependent $SU(2)$ matrix. However,
diagonalizing this matrix is not  
a covariant operation in this formalism.} 

In the rotated basis (denoted by the primed quantities)
the projector (\ref{012}) reads
\be312
\Gamma^4 \epsilon^{\prime A} 
+(\sigma^3)^A_{\ B}\, \epsilon^{\prime B} \ = \ 0\ ,
\ee
while $W$ simplifies to
\be311
W \ = \ \frac{2}{3}\, X^I\widehat P_I^{\prime 3}
\ .
\ee

The $\mu $ component of eq.\ (\ref{susyvar1})
leads to a first order differential equation 
which determines the $\mu$ dependence of $\epsilon^A$
\be314
2 \mu\, D_\mu \epsilon^A =   \epsilon^A\ \ ,
\ee
where we used (\ref{012}),(\ref{013}).
The compatibility of this equation with the projector
(\ref{012}) will be discussed after solving 
(\ref{susyvar2}) and (\ref{susyvar3})
which we turn to now.

Inserting the projector (\ref{012}) into  (\ref{susyvar2})
yields
\be320
\Big(\,\frac{1}{3} \mu\frac{d\phi^i}{d\mu} 
X^I+\partial^i X^I\Big)
\,\widehat P_{IB}^A\epsilon^B
+  X^I k_I^i\, \epsilon^A \ = \ 0\ .
\ee
Since $\widehat P_{IB}^A$ can be rotated into the
$\sigma^3$ direction the two terms of eq.\ (\ref{320})
have to vanish independently (unless all
supercharges are broken).
Thus with our choice of projector the isometries
of the vector multiplets cannot be gauged,
that is we have to demand $X^I k_I^i =0$.
The vanishing of the first term in (\ref{320})
imposes a first order differential equation 
for the scalar
fields 
\be014
%\beta^i\ \equiv\ 
\mu\,\frac{d\phi^i}{d\mu} \ = \ -3\,g^{ij}\partial_j\log
W~.
\ee

Finally, inserting the projector (\ref{012})
into the hyperino variation (\ref{susyvar3})
yields
\be016
\mu\,\frac{dq^u}{d\mu} \ = \ \frac{2}{3W^2} \
Tr(X^I\widehat 
P_I\,X^Jk_J^v J_v{}^u)~,
\ee
where we have used (\ref{013}) and the fact that the quaternionic 
complex structures are
given by
\be017
 (J_u{}^v)^A_{\ B} \ = \ 
g^{wv} (K_{uw})^A_{\ B}  \ = \ i\,V^{A\alpha}_u\,V^v_{B\alpha}~.
\ee  
Rewriting eq.\ (\ref{996}) as 
\be018
X^Ik_I^u (J)_u{}^v \ = \ g^{vu}\,D_u\,(X^I P_I)~,
\ee
can be used to recast (\ref{016}) in the form
\be019
\mu\,\frac{dq^u}{d\mu} \ = \ \frac{2}{3W^2} \,g^{uv}\,Tr\Big(X^I\widehat 
P_I\,D_v\,(X^J P_J)\Big) \ .
\ee
Eq.\ (\ref{019})
can also be written as a gradient flow 
 \be022
\mu\,\frac{dq^u}{d\mu}\ =\ -3\,g^{uv}\partial_v\log W 
\ee
with the same $W$ as in (\ref{014}), provided
\be033
Tr\Big(X^I\widehat 
P_I\,D_v\,(X^J P_J)\Big) = Tr\Big(X^I\widehat 
P_I\,D_v\,(X^J \widehat P_J)\Big)
\ee
holds.
This is a non-trivial constraint on the scalar 
function $\widehat{\gamma}_I(q)$ of eq.\ (\ref{031})
and is solved by
\be345
\widehat{\gamma}_{I}(q) \ = \ 1-\lambda_I\,(\det P_I)^{-1/2}\ ,
\quad \mbox{(no sum on $I$)} \ ,
\ee
where the $\lambda_I$ are arbitrary constants.  
For $\lambda_I = 0$ one has
$\widehat P_I =P_I$ and (\ref{022}) holds without any further condition.
However, eq.\ (\ref{345}) shows that also for non-trivial
$\widehat{\gamma}_I$ or in other words for a $\widehat P_I$ which
differs from $P_I$ the differential constraint on the hyper-scalars is
of the form (\ref{022}).  We will see in the next section that
non-trivial $\widehat{\gamma}_I$ are crucial in order to recover the
RG-flow of ref.\ \cite{040}.

Finally the compatibility of
eq.\ (\ref{012}) with (\ref{314})
imposes the additional condition
\be315
 D_\mu (W^{-1} X^I \widehat P^A_{I B}) =   0\ .
\ee
Using (\ref{678}), (\ref{310}) and (\ref{022}) this equation
is satisfied if $f_{IJ}^K = 0$, i.e.\ for
Abelian isometries.

It is known \cite{070,130,071} that for scalar
fields which obey eqs.\
(\ref{014}), (\ref{022}) the Einstein equations 
corresponding to the the action (\ref{020})
imply that the scalar potential $V$ has to take
the form
 \be043 
V \ = \  6\,\Big(\frac{3}{4}\, g^{MN}\partial_M W\partial_N W-W^2\Big) \ .
\ee
\remark{
Indeed, using the special geometry relations $G^{IJ}= \partial_i X^I
\partial_j X^J g^{ij} + \frac{2}{3} X^I X^J $ and the relation $W^2 \delta^A_B
= -4/9 {(X^I\hat{P}_I)^2}^A_B $ one can write the potential $V$ in the form
(\ref{043}) if the matrices $X^I \hat{P}_I $ and $\partial_i X^I \hat{P}_I $
commute. Clearly if $[\hat{P}_I , \hat{P}_J] = 0 $ for all $I,J$ (which
corresponds to Abelian gauged isometries ?), then this latter condition is
satisfied. However one might imagine a situation when some of the $\hat{P}_I$
do not commute. Now demanding $[\hat{P}_I , \hat{P}_J] = 0 $ implies that $
\partial_i X^I\hat{P}_I = a_i X^I \hat{P}_I $ with $a_i = \partial_i Log W $.
Since $\hat{P}_I $ are $SU(2)$ valued matrices, the maximum allowed subset
that do not commute is 3. Imagine an extreme case where $I= 0,1,2 $ and
$\hat{P}_I $ are proportional to $\sigma_1 , \sigma_2 , \sigma_3 $ for $I =
0,1,2 $ respectively (ie a 'maximally' non-commuting case). Then one finds
that the constraint implies the relation $\partial_i X^I = \lambda_i X^I $.
Such a constraint is incompatible with the special geometry condition $C_{IJK}
X^I X^J X^K = 1 $, with $C_{IJK} $ constants. In general one might have a
situation where a subset of the $\hat{P}_I $ do not commute, which we denote
by $\hat{P}_{I_1}$. In this case we would have differential constraints like
$\partial_i X^{I_1} = a_i X^{I_1} $ for a subset of the $X^I$ only. Whilst
this is not automatically incompatible with the special geometry constraint
mentioned earlier, it would seem to imply that the eigenvalues $a_i $ must only
depend on the coordinates $z^i $. But recalling the definition of $a_i $ this
places a strong constraint on the superpotential $W$ namely that $\partial_i
W = a_i(z) W$. It is difficult to see how this can be solved in general, other
than by requiring $\partial_i X^I = a_i(z) X^I $ as discussed above. Thus for
a generic special manifold parameterized by $X^I$ it is hard to see how one can
prove the relation eq (\ref{043})
 in cases where a (sub) set of $\hat{P}_I $ do not
commute. In any case, the example we will consider in detail later, will be
such that all $\hat{P}_I $ commute.
}
Indeed, inserting the special geometry relation $G^{IJ} =
\partial_iX^I\partial_jX^Jg^{ij} +\frac{2}{3}\,X^IX^J$, as well as
eqs.\ (\ref{996}),(\ref{310}), the relations (\ref{031}), (\ref{033})
and the definition
(\ref{013}) of $W$ into (\ref{997}) results in (\ref{043}).  Thus the
special form of the potential (\ref{043}) does not hold for an
arbitrary $N=2$ potential as given in (\ref{997}) but requires
precisely the same conditions that we needed to derive the flow
equations.\footnote{Strictly speaking eq.\ (\ref{043}) does not need
(\ref{310}) but already holds for the weaker condition
$[\partial_iX^I\widehat P_I,X^J\widehat P_J]=0\,$.  This relation is
certainly satisfied for (\ref{310}) but one could imagine a situation
where the $\widehat P_I$ not all commute but still satisfy
$[\partial_iX^I\widehat P_I,X^J\widehat P_J]=0\,$. However, this would
put strong constraints on $W$, following from the very special
geometry in 5-d which are not satisfied for standard choices of vector
scalar manifolds.}  
This can be viewed as a consistency check on our
procedure.

After this somewhat technical derivation let us summarize the results
and discuss the physical implications.  We solved the supersymmetric
variations (\ref{susyvar1})--(\ref{susyvar3}) in the background
(\ref{010}) and demanded four unbroken supersymmetries.  
This implies that only Abelian isometries of the
hypermultiplet geometry can be gauged,
i.e.\ $f_{IJ}^K = 0, X^I k_I^i=0$ with the further requirement 
$[\widehat P_I,\widehat P_J] = 0$. 
For the scalar fields a set of first
order differential equations follows.
Provided (\ref{033}) holds they 
can be written as gradient flow
equations  
\be023
\beta^M\ \equiv\ \mu\ \frac{d\Phi^M}{d\mu}\ = \ -3\,g^{MN}\,\partial_N\log
W~,
\ee
where
\be346
W \ = \ \frac{2}{3}\,
\Big(X^I X^J \widehat P_I^x \widehat P_J^x\Big)^{1/2} \
 = \ \frac{2}{3}\, X^I \widehat P_I^{\prime 3}~.
\ee
Eq.\ (\ref{023}) combines eqs.\ (\ref{014}) and (\ref{022}) while the
last equation in (\ref{346}) uses the fact that the $\widehat P_I$ can
always be chosen to point in the $\sigma^3$ direction.

The AdS/CFT
correspondence suggests  identifying the $\mu$-derivative of the scalar
fields with the $\beta$-function in the dual conformal field theory
\cite{351,352,040,120,121}.  The fixed points of the RG-flow occur for
$\partial_N W = 0$ which are also the extrema of the scalar potential
$V$ as can be seen from eq.\ (\ref{043}).  For $W|_{\partial W=0}\neq
0$ the extremum corresponds to an $AdS_5$ background with $W$ being
the cosmological constant.  $W|_{\p W=0} =0$ on the other hand
corresponds to a flat space-time background.

The nature of the fixed point is determined by
the derivatives of the $\beta$-functions
or more precisely by the eigenvalues of the matrix
\be779
\p_N \beta^M|_{\beta=0} \ = \ -3\,g^{MK}\,\frac{\p_N\p_K
W}{W}|_{\beta=0}
\ee
where we assume that the fixed point is non-singular, i.e.\ the metric
non-degenerate.  Negative eigenvalues correspond to fixed points that
are UV stable, while positive eigenvalues imply IR stable fixed
points.  A RG--flow configuration corresponds to a domain wall
interpolating between a UV and a IR point, whereas a Randall--Sundrum
type configuration interpolates between two IR fixed points, as we
will discuss in the conclusions.

Let us briefly discuss a few generic cases.
If there are no hypermultiplets in the spectrum
the Killing prepotentials are constants, i.e.
$P_I=\widehat P_I \equiv V_I =const.$
corresponding
to Fayet--Illiopoulos terms. 
In this case eq.\ (\ref{346}) implies 
\be333
W \ = \ \frac{2}{3}\,X^I(\phi)\, V_I\ .
\ee
This form of $W$
and the corresponding eqs.\ (\ref{043}), (\ref{023}) 
reproduce the results derived previously 
in \cite{020}. 
The very special geometry implies 
$\partial_i\partial_j W = \frac23  g_{ij} W$
and hence $\p_i \beta^j|_{\beta=0} = - 2 \delta_i^j$.
Thus all fixed points 
are necessarily ultraviolet \cite{110,020,090}. 
In other
words, neither RG-flows nor the RS scenario can be 
reproduced with only vector multiplets.\footnote{As
shown in \cite{090} this is also the case
if some of the vectors are dualized into tensor multiplets.}

If there are no vector multiplets but only 
hypermultiplets in the spectrum only the graviphoton
can be used as a gauge field and 
the superpotential reduces to
\be399
W \ = \ \frac{2}{3}\,X^0\,
\widehat P^{\prime 3}_0(q)\ .
\ee
Its second derivative does not have a fixed sign
 and thus $\partial_u\beta^v|_{\beta=0}$
can have positive and/or negative eigenvalues.

In the case that vector- and hypermultiplets are 
present 
the matrix $\p_N \beta^M|_{\beta=0} $
can have positive and negative eigenvalues.
Negative eigenvalues are necessarily present since
the submatrix $\partial_i\beta^j|_{\beta=0}$ always has
negative eigenvalues.
Thus any fixed point can be either a maximum
or a saddle point but not a  local minimum.
Positive eigenvalues can arise from 
the derivatives 
$\partial_u\beta^v|_{\beta=0}$ but also 
{}from mixed derivatives 
of the form $\partial_i\beta^v|_{\beta=0}$.
Finally, note that for
a superpotential that factorizes 
$W(q,\phi) = X(\phi) P(q)$
the mixed derivatives 
$\partial_i\beta^v|_{\beta=0}$
necessarily vanish.
However, the possibility of 
$\partial_u\beta^v|_{\beta=0}$ having positive 
eigenvalues still remains and thus 
non--trivial DW solutions are also possible
in this case. 

In ref.\ \cite{040} it was shown that whenever
the scalar fields obey gradient flow equations 
of the type (\ref{023}) the Einstein equations
of the Lagrangian (\ref{020}) imply a c-theorem
\cite{020,040,351,360} 
and that 
\be990 
C(\mu) \ = \ {C_0\over |W|^3}\ , \qquad C_0 = const.  
\ee 
is a natural candidate for the c-function.
This also holds in the setup here
and eqs.\ (\ref{023}) imply 
\be394
\mu
\frac{d}{d\mu}\, C \ = \ 
{1 \over |W|} g_{MN} \beta^M \beta^N \ > \ 0\ .
\ee
Thus $C$ is a monotonically increasing function of $\mu$ and
corresponds to the central charge at the conformal fixed points.

Before we turn to a specific example let us discuss the implications
of the condition (\ref{310}).  The vanishing commutator together with
(\ref{031}) implies that all $P_I(q)\,$ are proportional to each
other, i.e.\ $P_I(q)=\alpha_I(q) P(q)$.  {From} eq.\ (\ref{031}) we
learn that also
\be070
\widehat P_{IB}^A \ = \ \widehat\gamma_I\,\alpha_I\, P_{\ B}^A 
\equiv \gamma_I (q)\, P_{\ B}^A(q)\ 
\ee
holds.
This in turn says that only a $U(1)$ subgroup of
the $SU(2)_R$ is gauged with a gauge field
which is the ($q$-dependent) linear combination 
\be071
A_m \ =\ \sum_I \gamma_I(q)\,A_m^I \  ,
\ee 
Similarly, a $U(1)$ subgroup of the isometry group
is gauged -- albeit with a linear combination
of gauge fields that differs in the $q$-dependent coefficients.

For constant $\gamma_I$, we recover precisely the case considered in
refs.\ \cite{190,020,220,221} where only vector (and tensor)
multiplets are present.  With hypermultiplets we have the additional
possibilty that the linear combination of gauge fields is
$q$-dependent and thus can `rotate' along an RG-flow.  {From} the dual
$N=1$ field theory perspective this can be understood from the fact
that the non-anomalous $U(1)_R$ symmetry also changes along an RG-flow
\cite{151,152,801,800,040}.  Classically one has a $U(1)_R$ (often
denoted as the `standard' $U(1)_R$) which assigns zero $R$-charge to a
chiral superfield and $R$-charge $-1$ to the field strength $W_\alpha$
of the vector multiplet.  In addition, one generically has a flavour
symmetry $U(1)_K$ generated by the Konishi current which does
transform the chiral multiplets but leaves $W_\alpha$ invariant
\cite{151,152}.  Quantum mechanically the situation changes in that
the anomaly free $U(1)_R$ is in general a linear combination of the
standard $U(1)_R$ with the $U(1)_K$.  The coefficents of this linear
combination are related to the anomalous dimensions of certain
operators and hence they change along an RG-flow.  In the supergravity
description this fact is captured by eq.\ (\ref{071}).  We return to
this point in the next section.

%%%%%%%%%%%%%%%%%%%%%%%%%%%%%%%%%%%%%%%%%%%%%%%%%%%%%%%%%%

\section{Example of a BPS domain wall}

%%%%%%%%%%%%%%%%%%%%%%%%%%%%%%%%%%%%%%%%%%%%%%%%%%%%%%%%%%

In this section we discuss a specific BPS domain wall and show to what
extend the solution of ref.\ \cite{040} can be recovered.  In
\cite{040} a DW of gauged $N=8$ supergravity is given which
interpolates between two $AdS_5$ vacua of the scalar potential
$V$. One of the extrema preserves $N=8$ supersymmetry while the second
extrema only has $N=2$ supersymmetry and the interpolating kink
solution preserves $N=1$ supersymmetry.  In the AdS/CFT correspondence
this BPS-solution is identified with a RG-flow from an $N=4$ SCFT in
the UV to an $N=1$ SCFT in the IR which preserves $N=1$ supersymmetry
throughout the flow \cite{800}.

The gauge group of the $N=8$ supergravity is $SO(6)\sim SU(4)$ which
is identified with the R-symmetry of the $N=4$ SCFT.  This gauging
introduces a scalar potential $V$ which depends on the 42 scalars of
the $N=8$ gravitational multiplet spanning the coset
$E_{6(6)}/USp(8)$. 

In order to simplify the analysis the authors of \cite{040} decompose
the gauge group according to $SU(4) \to SU(2)_I\times SU(2)_G\times
U(1)_G$ and keep only $SU(2)_I$ singlets.  This corresponds to the
breaking $N=8\to N=4$ since 4 gravitini are projected out and
$SU(2)_G\times U(1)_G$ becomes the gauged R-symmetry of the $N=4$
supergravity.  In the scalar sector 11 scalars which are the singlets
of $SU(2)_I$ survive this projection.  It is shown that these 11
scalars span the coset
\be100
{\cal M} \ = \ {SO(5,2)\over SO(5)\times SO(2)} 
\times SO(1,1) \ ,
\ee
which is precisely the scalar manifold of
two $N=4$ tensor multiplets 
coupled to $N=4$ supergravity \cite{409}.
An $N=4$ tensor multiplet contains an antisymmetric 
tensor, four fermions and five scalars 
while the $N=4$ gravitational multiplet contains 
the gravition, four gravitini, six graviphotons,
four fermions and one scalar. This scalar
spans the $SO(1,1)$ component of ${\cal M}$.

In order to make contact with the previous section 
we need to do a further truncation to $N=2$ 
along the lines of ref.\ \cite{408}.
This can be done by  decomposing
the gauge group further and again projecting onto 
invariant states.
More precisely we decompose 
\be101
SU(2)_G\times U(1)_G \ \to \ U(1)_3\times U(1)_G
\ee
where $U(1)_3$ is the $U(1)$ generated by $\sigma^3$ 
inside $SU(2)_G$.
We only keep states which are invariant under the
diagonal subgroup of $U(1)_3\times U(1)_G$.
This leaves one $U(1)_R$
(with the other combination of charges) intact.

The 8 gravitini of $N=8$ supergravity transform as a ${\bf 4}\oplus{\bf
\bar 4}$ of $SU(4)$.  In the decomposition $SU(4)\to SU(2)_I\times
SU(2)_G\times U(1)_G$ the 4 $SU(2)_I$ invariant gravitini transform
according to the ${\bf 2}_{-1/2}\oplus {\bf \bar 2}_{1/2} $ of
$SU(2)_G\times U(1)_G$. The $U(1)_3\times U(1)_G$ invariance projects
out two more gravitini leaving two complex conjugate gravitini
transforming under $U(1)_R$.  The rest of the $N=4$ spectrum can be
similiarly truncated. Out of the four vectors of $SU(2)_G\times
U(1)_G$ two Abelian vectors of $U(1)_3\times U(1)_G$ survive.  The two
tensors are both projected out while out of the 11 scalars 5 survive.
The 5 scalars in the tensor multiplet reside in the representation
$
{\bf 3}_{1} \oplus{\bf 1}_{2} \oplus {\bf 1}_0
$
while the second tensor multiplet carries the complex conjugate
representation.  Thus after projection one is left with two singlets
and two $U(1)_R$ charged scalars.  The 5th scalar comes out of the
gravitational multiplet and is a singlet of $SU(2)_G\times U(1)_G$ and
thus also of $U(1)_R$.  
Out of the 12 fermions which reside in 
the $ {\bf 2}_{-3/2} \oplus{\bf 2}_{+3/2} 
\oplus 2\times ({\bf 2}_{-1/2} \oplus{\bf 2}_{+1/2})$
of $SU(2)_G\times U(1)_G$ four survive the projection.
The surviving states fit
precisely into one gravitational multiplet, one vector multiplet $V$
and one hypermultiplet $H$ of $N=2$ supergravity.  The scalar in $V$
is neutral under $U(1)_R$ while the hypermultiplet hosts the two
neutral and the two charged scalars.  The two neutral scalars of the
hypermultiplet can be identified with the dilaton and axion of type
IIB supergravity.

The AdS/CFT correspondence relates these five scalars to gauge
invariant operators in the dual CFT.  The UV theory is an $N=4$ SCFT
with Yang-Mills gauge group $G=SU(n)$. Written in terms of $N=1$
superfields this theory has one vector multiplet, three chiral
multiplet in the adjoint representation of $G$ and a superpotential
$W= Tr A_1[A_2,A_3]$. The RG-flow is induced by adding the operator $m
Tr A_3^2$ to $W$ \cite{800,040,801}.  The non--anomalous $U(1)$
symmetry discussed at the end of section 3 is a linear combination of
the $U(1)_R$ at $m=0$ which assign zero R-charge to all three
superfields $A$ and the Konishi $U(1)_K$ symmetry which assigns $A_3$
a $U(1)_K$ charge $-1$.  In the dual supergravity the dilaton and
axion play the role of the gauge coupling and the $\theta$-angle,
respectively. The charged scalar $C$ couples to the operator $Tr
A_3^2$ and the five-dimensional vector couples to the Konishi current.

The resulting
scalar manifold of the supergravity 
can be derived by truncating
${\cal M }$ given in eq.\ (\ref{100}).
The $SO(1,1)$ factor of ${\cal M }$ survives
the projection since its scalar is invariant.
This component is a one-dimensional 
very special K\"ahler manifold
characterized by \cite{220}
\be106
{\cal V} \ = \ X^0(X^1)^2 \ = \ 1\ .
\ee

The coset space ${SO(5,2)\over SO(5)\times SO(2)}$
is a K\"ahler manifold with a K\"ahler potential
\be102
K \ = \ -\frac12\ln\Big[(S+\bar S)(T+\bar T) - \frac12
\sum_{i=1}^3 (C+\bar C)_i^2\Big]~.
\ee
The three $C_i$ are the triplet while $S$ and $T$
are singlets of $SU(2)_G$.
In addition one has 
$SL(2,{\bf R})\times SL(2,{\bf R})$ acting as fractional
linear transformations on all fields.
One $SL(2,{\bf R})$ is the symmetry associated with 
the dilaton while the other  $SL(2,{\bf R})$
hosts the $U(1)_G$ as its compact subgroup \cite{040}.
Thus projecting onto $U(1)_3\times U(1)_G$
invariant fields leaves $S$ (or $T$) and one
of the three $C_i$ which we denote by $C$ henceforth.
The K\"ahler potential becomes
\be103
K \ = \ -\frac12\ln\Big[(S+\bar S)- \frac12 (C+\bar C)^2\Big]\ ,
\ee
which is the K\"ahler potential of the coset space 
${SU(2,1)\over U(2)}$. This is indeed a quaternionic 
manifold known as the ``universal hypermultiplet''
\cite{412}.
Hence, the combined 
scalar manifold of the $N=2$ supergravity
is found to be
\be107
{\cal M} \ = \ SO(1,1) \times {SU(2,1)\over U(2)}\ .
\ee

In the following we use a more convenient
parameterization by shifting
$S\to S +\frac12 C^2$ which results in
\be104
K \ = \ -\frac12\ln\Big[(S+\bar S)- C \bar C\Big]\ ,
\ee
In these variables the $U(1)_R$ acts as
\be105
C \ \to \ e^{i\theta} C ,\qquad S \ \to \ S~ .
\ee

The next step is to gauge the isometry 
(\ref{105}).\footnote{In ref.\ \cite{190}
the same quaternionic geometry was considered
but a different isometry corresponding
the shift $S\to S+\alpha, C\to C$ was gauged.
This leads to a different potential which does
not correspond to a RG-type domain wall.}
To do so we need to briefly recall the quaternionic
quantities of ${SU(2,1)\over U(2)}$ \cite{412}.
We use as quaternionic coordinates
$q^u = (S,\bar S,C,\bar C)$.
In these coordinates the $SU(2)$ connection reads
\bea\label{112}
\omega_S &=& \frac14 e^{2K} \sigma^3\ ,\qquad
\omega_{\bar S} \ = \ -\frac14 e^{2K} \sigma^3\ ,
\\
\omega_C &=&\left(\ba{cc} -\frac14 e^{2K}\bar C &-e^K\\
           0&\frac14 e^{2K}\bar C\ea\right)\ ,\qquad
\omega_{\bar C} \ = \ \left(\ba{cc} \frac14 e^{2K} C &0\\
    e^{K}  &-\frac14 e^{2K} C\ea\right)\ .\nonumber 
\eea
The matrix of hyper K\"ahlerforms $(K_{uv})_{\ B}^A$ 
%=K_{uv}^x (\sigma^x)_A^B$ 
is given by
 \bea\label{113}
K_{S\bar S} &=& -\frac12 e^{4K} \sigma^3\ ,\qquad
K_{C\bar C} \ = \ \left(\ba{cc} 
\frac12 e^{2K}(1-e^{2K}C \bar C)& -e^{3K}C \\
   -e^{3K}\bar C&-\frac12 e^{2K}(1-e^{2K}C \bar C)
\ea\right)\nonumber \\
K_{S\bar C} &=& 
\left(\ba{cc} \frac12e^{4K}C &0\\
       e^{3K}     &-\frac12 e^{4K}C\ea\right)\ ,\qquad
K_{\bar S C} \ = \ 
\left(\ba{cc}- \frac12e^{4K} \bar C& -e^{3K}\\ 0&\frac12 e^{4K}\bar C\ea\right)
\ ,
\eea
while all other components are zero.
The Killing vector for the symmetry (\ref{105})
is
\be114
k^u \ = \ \left(0,0,i\,C,-i\,\bar C\right)\ .
\ee
Using (\ref{112})--(\ref{114}) the solution of eq.\
(\ref{996}) is found to be
\be115
P^A_{\ B} \ = \ \frac{i}{2}\,
\left(\ba{cc} 1- e^{2K}C\bar C & 
-2 e^K\,C \\ -2e^K\,\bar C &  -(1- e^{2K} C\bar C)
 \ea\right)\ .
\ee

As we stated above the $U(1)_R$ is a linear combination
of the $U(1)_3$ and the $U(1)_G$
and therefore both gauge fields appear
in the covariant derivatives.
Using (\ref{071}) we allow this linear
combination to be $q$-dependent and 
from (\ref{070}) we infer that 
the $\widehat P$ obey
\be116
\widehat P_0 \ = \ \gamma_0 P\ ,\qquad 
\widehat P_1 \ = \ \gamma_1 P\ ,
\ee
where $P$ is given by (\ref{115})
and the $\gamma_I$ satisfy (\ref{345}).
Inserted into (\ref{346}) the resulting superpotential is
\be117
W \ = \  \frac{1}{3}\,(\gamma_0X^0+\gamma_1X^1)\
\frac{S+\bar S}{S+\bar S-C\bar C}~.
\ee
Using the constraint (\ref{106}) 
and introducing new variables 
\be130
X^1 \ \equiv \ \rho^{-2}\quad,\qquad 
\frac{C\bar C}{S+\bar S} \ \equiv \ \tanh^2(\chi)~,
\ee
yields
\be131
W \ = \
\frac{1}{3\rho^2}\,(\gamma_0\,\rho^6+\gamma_1)
\,\cosh^2(\chi)~.
\ee  
For the choice 
\be132
\gamma_0 \ = \ \frac32(2\tanh^2(\chi) - 1)\quad ,
\qquad
\gamma_1 \ = \ -3~,
\ee
one obtains
\be134
W \ = \
\frac{1}{4\rho^2}\,\Big[(\rho^6-2)\cosh(2\chi)-
(3 \rho^6 +2)\Big] \ ,
\ee  
which precisely coincides with the superpotential of ref.\ 
\cite{040}.
The RG-flow governed by this superpotential  
has a UV fixed point at $\rho=1 ,\chi=0$
and an IR fixed point at $\rho^6=2 ,2\chi=\log 3$.
It is important to note that with constant
$\gamma_0$, $W$ factorizes as can be seen from
(\ref{131})
and it is not possible to recover
(\ref{134}). 
Thus it is crucial to allow for 
$q$-dependent $\gamma_0$ and this is the main
motivation for introducing $\widehat \gamma_I$
in eq.\ (\ref{031}).  The specific $\gamma_0$
of (\ref{132}) indeed
satisfies the constraint (\ref{345})
for $\alpha_0 = 3/2,~\alpha_1 = -3$.

Finally let us note that the dilaton $S$ 
automatically stays constant along this flow.
Using (\ref{023}), (\ref{103}), (\ref{117}) and (\ref{132})
one finds
\be133
\mu {d\over d\mu} S \ = \ -{3\over W} 
(g^{S\bar S} \partial_{\bar S} W 
+  g^{S\bar C} \partial_{\bar C} W) \ = \ 0 \ .
\ee
This can be viewed as a consistency check of our solution.
Note that for the universal hypermultiplet a constant dilaton along
the RG-trajectory is not a special feature of (\ref{134})
but holds for any superpotential $W$ 
which is a function of $\frac{C\bar C}{S+\bar S}\,$
only as can be easily verified from eq.\ (\ref{133}).

\section{Conclusions}

In this paper we derived $N=1$ BPS domain wall solutions of gauged
five-dimensional $N=2$ supergravity.  
Our main result are the
supersymmetric flow equations (\ref{023}), (\ref{346}) which include
scalars from vector and hypermultiplets.  The presence of charged
hypermultiplets turns out to be crucial in recovering IR-fixed points
in RG-flows of a dual (perturbed) superconformal field theory.
In order to recover the specific flow of 
ref.\ \cite{040} it is necessary to modify 
the standard gauging of the R-symmetry.
The validity of this modification 
remains to be rigorously proven.
However, the fact that we do recover the flow of
\cite{040} can also be viewed as a consistency
check.

The necessity of IR fixed points in order to construct
a smooth supersymmetric domain-wall solution
of the Randall-Sundrum type has been stressed 
in ref.\ \cite{090}.
Let us briefly recall the argument.
It is
convenient to first change coordinates and replace 
the $\mu$ of the Ansatz (\ref{010})
by
\be500
\mu \ = \ e^{A(z)}\ ,\qquad W \ = \ \partial_z A(z)\ .
\ee
In these coordinates the metric  (\ref{010})
reads
\be510
ds^2 \ = \ e^{2A(z)} \Big( -dt^2 + d\vec x^2 \Big) + dz^2\ ,
\ee
which is the metric of $AdS_5$ for $A=\pm k z, k= const.$.
In these coordinates the 
UV fixed point ($\mu\to\infty$) of an RG-flow
is located at $z\to\infty$ where $A\sim z\to +\infty$
while the IR fixed point ($\mu=0$)
sits at $z\to-\infty$ where $A\sim z\to -\infty$.
The DW solution 
interpolates between the two asymptotic regions
at $z=\pm\infty$.

A smooth DW solution corresponding to the RS-setup needs to have a
different asymptotic behavior. In that case one has a ${\bf Z_2}$
symmetric solution with $A\to -k|z|$ for $z\to \pm\infty$ \cite{050}.
That is one has a decreasing warp factor at both ends $z\to\pm\infty$
or in the language of the RG-flow a DW solution connecting two
IR-fixed points \cite{090}.  Obviously, such a solution cannot be
interpreted as an RG-flow. $A(z)$ has at least one maximum where $W =
\partial A(z) =0$.  At that point the $\beta$-functions of eqs.\
(\ref{023}) as well as the c-function of (\ref{990}) become singular \cite{888}.
However, even if such DW solutions do not make sense as RG-flows there
is no obvious reason why they should not exist. The previous no-go
theorems merely stated that they cannot be found with only non-trivial
vector and tensor multiplets.  Adding charged hypermultiplets changes
the story and we are optimistic that supersymmetric RS-domain walls do
exist \cite{007}.  They should be smooth generalizations of the
constructions presented in refs.\ \cite{271,272,273,274,275,320}.

Finally, IR-fixed points have also been recently studied
by wrapping M5-branes on a Riemann surface of constant
negative curvature in the presence of a non-trivial gauge field
\cite{999}.  Following this procedure, one obtains after
compactification to 5 dimensions a 3-brane solution with an $AdS_5$
vacuum, which is IR attractive (near the AdS horizon). 
But in the UV
limit this solution decompactifies into $AdS_7$, 
i.e.\ is singular
from the 5-dimensional perspective. 
It would be interesting to study this situation within
the formalism of this paper.

\bigskip

{\bf Acknowledgments}

The work of K.B.\ is supported by a Heisenberg grant of the DFG.
The work of C.H.\ and J.L.\ is 
supported by
GIF -- the German--Israeli
Foundation for Scientific Research and the 
DAAD -- the German Academic Exchange Service.
The work of S.T. was funded by the British-German
ARC programme.

We thank H.\ G\"unther,
R.\ Kallosh, T.\ Mohaupt,
S.\ Pokorski, L.\ Randall,  L.\ Thorlacius,
D.\ Waldram, N.\ Warner  and especially
A.\ Ceresole and G.\ Dall'Agata
for discussions and correspondence.
C.H.\ thanks P.\ di Vecchia for 
an invitation to the Nils Bohr Institute
at an early stage of this project.
J.L.\ would like to thank the Erwin Schr\"odinger
Institute for hospitality and financial support
during part of this work. 

%%%%%%%%%%%%%%%%%%%%%%%%%%%%%%%%%%%%%%%%%%%%%%%%%%%%%%%%%%%%%%

% ---- Bibliography ----

% \nocite{*}                   %this uses *everything* in the .bib file
% \bibliography{mar00}          %or whatever your .bib file is

\begin{thebibliography}{10}

\bibitem{401}
O.~Aharony, S.~S.~Gubser, J.~Maldacena, H.~Ooguri and Y.~Oz,
``Large N field theories, string theory and gravity'',
Phys.\ Rept.\  {\bf 323} (2000) 183,
\href{http://xxx.lanl.gov/abs/hep-th/9905111}{{\tt hep-th/9905111}}.

\bibitem{402}
J.~Maldacena,
``The large N limit of superconformal field theories and supergravity'',
Adv.\ Theor.\ Math.\ Phys.\  {\bf 2} (1998) 231,
\href{http://xxx.lanl.gov/abs/hep-th/9711200}{{\tt hep-th/9711200}}.

\bibitem{403}
S.~Kachru and E.~Silverstein,
``4d conformal theories and strings on orbifolds'',
Phys.\ Rev.\ Lett.\  {\bf 80} (1998) 4855,
\href{http://xxx.lanl.gov/abs/hep-th/9802183}{{\tt hep-th/9802183}}.

\bibitem{411}
Y.~Oz and J.~Terning,
``Orbifolds of AdS(5) x S(5) and 4d conformal field theories'',
Nucl.\ Phys.\  {\bf B532} (1998) 163,
\href{http://xxx.lanl.gov/abs/hep-th/9803167}{{\tt hep-th/9803167}}.

\bibitem{404}
I.~R.~Klebanov and E.~Witten,
``Superconformal field theory on threebranes at a Calabi-Yau  singularity'',
Nucl.\ Phys.\  {\bf B536} (1998) 199,
\href{http://xxx.lanl.gov/abs/hep-th/9807080}{{\tt hep-th/9807080}}; \\
``AdS/CFT correspondence and symmetry breaking'',
Nucl.\ Phys.\  {\bf B556} (1999) 89,
\href{http://xxx.lanl.gov/abs/hep-th/9905104}{{\tt hep-th/9905104}}.

\bibitem{456}
B.S.\ Acharya, J.M.\ Figueroa-O'Farrill, C.M.\ Hull
and B.\ Spence,
``Branes at conical singularities and holography'',
Adv.\ Theor.\ Math.\ Phys. {\bf 2} (1998) 1249,
\href{http://xxx.lanl.gov/abs/hep-th/9808014}{{\tt hep-th/9808014}}.

\bibitem{405}
D.~R.~Morrison and M.~R.~Plesser,
``Non-spherical horizons. I'',
Adv.\ Theor.\ Math.\ Phys.\  {\bf 3} (1999) 1,
\href{http://xxx.lanl.gov/abs/hep-th/9810201}{{\tt hep-th/9810201}}.

\bibitem{351}
L.~Girardello, M.~Petrini, M.~Porrati, and A.~Zaffaroni, ``Novel Local CFT
and
exact results on perturbation of $N=4$ 
Super Yang-Mills dynamics from AdS Dynamics'',
JHEP {\bf 12} (1998) 022,
  \href{http://xxx.lanl.gov/abs/hep-th/9810126}{{\tt hep-th/9810126}}.

\bibitem{352}
J.\ Distler and F.\ Zamora, 
``Nonsupersymmetric Conformal Field Theories
from Stable Anti- de Sitter Spaces'',
 Adv.\ Theor.\ Math.\ Phys.\  {\bf 2} (1999) 1405,
 \href{http://xxx.lanl.gov/abs/hep-th/9810206}{{\tt hep-th/9810206}}.

\bibitem{040}
D.~Z. Freedman, S.~S. Gubser, K.~Pilch, and N.~P. Warner,
``Renormalization
  group flows from holography-supersymmetry and a c-theorem'',
  \href{http://xxx.lanl.gov/abs/hep-th/9904017}{{\tt hep-th/9904017}}.

\bibitem{041}
A.~Khavaev, K.~Pilch and N.~P.~Warner,
``New vacua of gauged N = 8 supergravity in five dimensions'',
\href{http://xxx.lanl.gov/abs/hep-th/9812035}{{\tt hep-th/9812035}}.

\bibitem{120}
J.~de~Boer, E.~Verlinde, and H.~Verlinde, ``On the holographic
renormalization
  group'', \href{http://xxx.lanl.gov/abs/hep-th/9912012}{{\tt
hep-th/9912012}}.

\bibitem{121}
C.~Schmidhuber,
``AdS-flows and Weyl gravity'',
Nucl.\ Phys.\  {\bf B580} (2000) 121,
\href{http://xxx.lanl.gov/abs/hep-th/9912155}{{\tt hep-th/9912155}}.

\bibitem{406}
E.~Witten,
``Anti-de Sitter space and holography'',
Adv.\ Theor.\ Math.\ Phys.\  {\bf 2} (1998) 253,
\href{http://xxx.lanl.gov/abs/hep-th/9802150}{{\tt hep-th/9802150}}.

\bibitem{407}
A.~W.~Peet and J.~Polchinski,
``UV/IR relations in AdS dynamics'',
Phys.\ Rev.\  {\bf D59} (1999) 065011,
\href{http://xxx.lanl.gov/abs/hep-th/9809022}{{\tt hep-th/9809022}}.

\bibitem{410}
K.~Pilch and N.~P.~Warner,
``N = 2 supersymmetric RG flows and the IIB dilaton'',
\href{http://xxx.lanl.gov/abs/hep-th/0004063}{{\tt hep-th/0004063}}.

\bibitem{408}
K.~Pilch and N.~P.~Warner,
``N = 1 supersymmetric renormalization group flows from IIB supergravity'',
\href{http://xxx.lanl.gov/abs/hep-th/0006066}{{\tt hep-th/0006066}}.

\bibitem{050}
L.~Randall and R.~Sundrum, ``An alternative to compactification'', 
Phys.\  Rev.\ Lett.\ {\bf 83} (1999) 4690,
  \href{http://xxx.lanl.gov/abs/hep-th/9906064}{{\tt hep-th/9906064}}; \\
``A large mass hierarchy from a small extra
  dimension'',  Phys.\ Rev.\ Lett.\ {\bf 83} (1999) 3370,
  \href{http://xxx.lanl.gov/abs/hep-ph/9905221}{{\tt hep-ph/9905221}}.

\bibitem{271}
R.~Altendorfer, J.~Bagger and D.~Nemeschansky,
``Supersymmetric Randall-Sundrum scenario'',
\href{http://xxx.lanl.gov/abs/hep-th/0003117}{{\tt
  hep-th/0003117}}.

\bibitem{272}
T.~Gherghetta and A.~Pomarol,
``Bulk fields and supersymmetry in a slice of AdS'',
\href{http://xxx.lanl.gov/abs/hep-ph/0003129}{{\tt
  hep-ph/0003129}}.

\bibitem{273}
A.~Falkowski, Z.~Lalak and S.~Pokorski,
``Supersymmetrizing branes with bulk in five-dimensional supergravity'',
\href{http://xxx.lanl.gov/abs/hep-th/0004093}{{\tt
  hep-th/0004093}}.

\bibitem{274}
E.~Bergshoeff, R.~Kallosh and A.~Van Proeyen,
``Supersymmetry in singular spaces'',
\href{http://xxx.lanl.gov/abs/hep-th/0007044}{{\tt
  hep-th/0007044}}.

\bibitem{275}
M.~J.~Duff, J.~T.~Liu and K.~S.~Stelle,
``A supersymmetric type IIB Randall-Sundrum realization'',
\href{http://xxx.lanl.gov/abs/hep-th/0007120}{{\tt
  hep-th/0007120}}.

\bibitem{320}
M.~Cvetic, H.~Lu, and C.~N. Pope, ``Domain walls with localised gravity and domain-wall/QFT correspondence'', \href{http://xxx.lanl.gov/abs/hep-th/0007209}{{\tt
  hep-th/0007209}}.

\bibitem{090}
R.~Kallosh and A.~Linde, ``Supersymmetry and the brane world'',  
JHEP {\bf 02} (2000) 005, 
\href{http://xxx.lanl.gov/abs/hep-th/0001071}{{\tt hep-th/0001071}}.

\bibitem{110}
K.~Behrndt and M.~Cveti{\v c}, ``Anti-deSitter vacua of gauged
supergravities
  with 8 supercharges'',
Phys.\ Rev.\  {\bf D61} (2000) 101901,
\href{http://xxx.lanl.gov/abs/hep-th/0001159}{{\tt
  hep-th/0001159}}.

\bibitem{190}
A.~Lukas, B.~A. Ovrut, K.~S. Stelle, and D.~Waldram, ``Heterotic
{M}-theory in
  five dimensions'', Nucl.\ Phys.\ {\bf B552} (1999) 246,
  \href{http://xxx.lanl.gov/abs/hep-th/9806051}{{\tt hep-th/9806051}};\\
``The Universe as a Domain Wall'',  Phys.\ Rev.\ {\bf D59}
(1999) 086001,
 \href{http://xxx.lanl.gov/abs/hep-th/9803235}{{\tt hep-th/9803235}}.

\bibitem{250}
J.~Ellis, Z.~Lalak, and W.~Pokorski, ``Five-dimensional gauged
supergravity and
  supersymmetry breaking in {M}-theory'', Nucl.\ Phys.\ {\bf B559}
(1999) 71, 
\href{http://xxx.lanl.gov/abs/hep-th/9811133}{{\tt hep-th/9811133}}.

\bibitem{020}
K.~Behrndt, ``Domain walls of {D} = 5 supergravity and fixed points of 
{N}= 1 super Yang-Mills'', 
 Nucl.\ Phys.\ {\bf B573} (2000) 127, 
\href{http://xxx.lanl.gov/abs/hep-th/9907070}{{\tt
  hep-th/9907070}}.

\bibitem{010}
K.~Behrndt and M.~Cveti{\v c}, ``Supersymmetric domain wall world from
{D=5}
  simple gauged supergravity'',
Phys.\ Lett.\ {\bf B475} (2000) 253,
  \href{http://xxx.lanl.gov/abs/hep-th/9909058}{{\tt hep-th/9909058}}.

\bibitem{220}
M.~Gunaydin and M.~Zagermann, ``The gauging of five-dimensional, {N=2}
  {M}axwell-{E}instein supergravity theories coupled to tensor
multiplets'',
 Nucl.\ Phys.\ {\bf B572} (2000) 131, 
  \href{http://xxx.lanl.gov/abs/hep-th/9912027}{{\tt hep-th/9912027}}.

\bibitem{210}
M.~Gunaydin and M.~Zagermann, ``The vacua of 5d, N = 2 gauged
  {Yang-Mills/Einstein}/tensor supergravity: {A}belian case'',
Phys.\ Rev.\  {\bf D62} (2000) 044028,
  \href{http://xxx.lanl.gov/abs/hep-th/0002228}{{\tt hep-th/0002228}}.

\bibitem{221}
A.~Ceresole and G.~Dall'Agata,
``General matter coupled N = 2, D = 5 gauged supergravity'',
\href{http://xxx.lanl.gov/abs/hep-th/0004111}{{\tt hep-th/0004111}}.

\bibitem{201}
M.~Cveti{\v c}, S.~Griffies, and S.-J. Rey, ``Static domain walls in {N=1}
  supergravity'',  Nucl.\ Phys.\  {\bf B381} (1992) 301--328,
  \href{http://xxx.lanl.gov/abs/hep-th/9201007}{{\tt hep-th/9201007}};
M.~Cveti{\v c} and H.~H. Soleng, ``Supergravity domain walls'',  
Phys.\  Rept.\  {\bf 282} (1997) 159,
  \href{http://xxx.lanl.gov/abs/hep-th/9604090}{{\tt hep-th/9604090}}.

\bibitem{200}
L.~Andrianopoli, M.~Bertolini, A.~Ceresole, 
R.~D'Auria, S.~Ferrara and P.~Fre, 
``General matter coupled {N=2} supergravity'',
Nucl.\ Phys.\ {\bf B476} (1996) 397--417,
\href{http://xxx.lanl.gov/abs/hep-th/9603004}{{\tt hep-th/9603004}};\\
L.~Andrianopoli, M.~Bertolini, A.~Ceresole, R.~D'Auria, S.~Ferrara, P.~Fre 
and T.~Magri,
``N = 2 supergravity and N = 2 super Yang-Mills theory on general scalar  
manifolds: Symplectic covariance, gaugings and the momentum map'',
J.\ Geom.\ Phys.\  {\bf 23} (1997) 111,
\href{http://xxx.lanl.gov/abs/hep-th/9605032}{{\tt hep-th/9605032}}.

  
\bibitem{409}
M.~Gunaydin, L.~J.~Romans and N.~P.~Warner,
``Compact And Noncompact Gauged Supergravity Theories In Five-Dimensions'',
Nucl.\ Phys.\  {\bf B272} (1986) 598.

\bibitem{080}
M.~Gunaydin, G.~Sierra, and P.~K. Townsend, ``Gauging the {D=5}
  {M}axwell-{E}instein supergravity theories: More on {J}ordan algebras'',
  Nucl.\ Phys.\ {\bf B253} (1985) 573.

\bibitem{070}
P.~K. Townsend, ``Positive energy and the scalar potential in higher
  dimensional (super)gravity theories'',  Phys.\ Lett.\ {\bf B148}
(1984) 55.

\bibitem{130}
K.~Skenderis and P.~K. Townsend, ``Gravitational stability and
  renormalization-group flow'',  Phys.\ Lett.\ {\bf B468} (1999) 46,
  \href{http://xxx.lanl.gov/abs/hep-th/9909070}{{\tt hep-th/9909070}}.

\bibitem{071}
A.~Chamblin and G.~W.~Gibbons,
``Nonlinear supergravity on a brane without compactification'',
Phys.\ Rev.\ Lett.\  {\bf 84} (2000) 1090,
\href{http://xxx.lanl.gov/abs/hep-th/9909130}{{\tt hep-th/9909130}}.

\bibitem{360}
E.~Alvarez and C.~Gomez, ``Geometric holography, the renormalization group
and
  the c-theorem'',  Nucl.\ Phys.\ {\bf B541} (1999) 441,
  \href{http://xxx.lanl.gov/abs/hep-th/9807226}{{\tt hep-th/9807226}}.

\bibitem{151}
N.~Seiberg,
``Electric - magnetic duality in supersymmetric 
non-Abelian gauge theories'',
Nucl.\ Phys.\  {\bf B435}, 129 (1995)
\href{http://xxx.lanl.gov/abs/hep-th/9411149}{{\tt hep-th/9411149}}.

\bibitem{152}
I.~I.~Kogan, M.~Shifman and A.~Vainshtein,
``Matching conditions and duality in {N=1 SUSY} gauge theories in the 
conformal window'',
Phys.\ Rev.\  {\bf D53}, 4526 (1996),
\href{http://xxx.lanl.gov/abs/hep-th/9507170}{{\tt hep-th/9507170}}.

\bibitem{801}
R.\ Leigh and M.\ Strassler,
``Exactly marginal operators and duality in four-dimensional N=1
supersymmetric gauge theory'', 
Nucl.\ Phys.\  {\bf B447} (1995) 95,
\href{http://xxx.lanl.gov/abs/hep-th/9503121}{{\tt hep-th/9503121}}.

\bibitem{800}
A.~Karch, D.~L\"ust and A.~Miemiec,
``New N = 1 superconformal field theories and their supergravity
description'', Phys.\ Lett.\  {\bf B454} (1999) 265,
\href{http://xxx.lanl.gov/abs/hep-th/9901041}{{\tt hep-th/9901041}}.

\bibitem{412}
S.~Ferrara and S.~Sabharwal,
``Quaternionic Manifolds For Type {II}
 Superstring Vacua Of Calabi-Yau Spaces'',
Nucl.\ Phys.\  {\bf B332} (1990) 317.

\bibitem{888}
G.~W.~Gibbons and N.~D.~Lambert,
``Domain walls and solitons in odd dimensions'',
{\tt hep-th/0003197}.

\bibitem{007}
K.~Behrndt, C.~Herrmann, J.~Louis and S.~Thomas, in preparation.

\bibitem{999}
J.~Maldacena and C.~Nunez, ``Supergravity description of field theories 
on curved manifolds and a no  go theorem'',
{\tt hep-th/0007018}.

\end{thebibliography}
% \bibliographystyle{unsrt}   %if you use utphys.bst
%\href{http://xxx.lanl.gov/abs/hep-th/}{{\tt hep-th/}}.

\providecommand{\href}[2]{#2}
\begingroup
%\raggedright
\endgroup

\end{document}